# Exceptionally large anomalous Hall effect due to anticrossing of spin-split bands in the antiferromagnetic half-Heusler compound TbPtBi

Yanglin Zhu[1], Bahadur Singh[2,*], Yu Wang[1], Cheng-Yi Huang[3], Wei-Chi Chiu[2], Baokai Wang[2], David Graf [5], Yubo Zhang[5], Hsin Lin[3], Jianwei Sun[5], Arun Bansil[2,*] and Zhiqiang Mao[1,*]

[1] Department of Physics, Pennsylvania State University, University Park, PA, 16802

[2] Department of Physics, Northeastern University, Boston, USA, 02115

[3] Institute of Physics, Academia Sinica, Taipei 11529, Taiwan

[4] National High Magnetic Field Laboratory, Tallahassee, FL, 32310

[5] Physics and Engineering physics department, Tulane University, New Orleans, LA, 70118

## Abstract

We have investigated magnetotransport properties and the topological electronic structure of the half-Heusler compound TbPtBi. Our experiments reveal an exceptionally large anomalous Hall effect (AHE) in the canted antiferromagnetic state of TbPtBi with the anomalous Hall angle (AHA) reaching ~0.68-0.76, which is a few times larger than the previously reported record in GdPtBi. First-principles electronic structure and the associated anomalous Hall conductivity were computed in order to interpret the experimental results. Our analysis shows that the AHE in TbPtBi does not originate from the Weyl points but that it is driven by the large net Berry curvature produced by the anticrossing of spin-split bands near the Fermi level in TbPtBi.

*emails: bahadursingh24@gmail.com; zim1@psu.edu; ar.bansil@northeastern.edu

The ordinary Hall effect in non-magnetic metals and semiconductors arises when oppositely charged carriers are separated via the Lorentz force in the presence of an external magnetic field. In a magnetic system, however, the anomalous Hall effect (AHE) can occur, which does not require an external magnetic field. The AHE is not only of fundamental scientific interest, but it also carries promise for technological applications (e.g. Hall sensors). Three different mechanisms for the AHE have been identified through studies of a variety of materials [1-4]: (i) An extrinsic effect due to spin-dependent scattering involving skew scattering and side jump [2,3,5]; (ii) An intrinsic effect originating from the Berry curvature in real space [6-10], which usually occurs in antiferromagnetic (AFM) systems with spin chirality [11-17]; and, (iii) an intrinsic effect induced by the Berry curvature in momentum space [18-24], which is predicted in magnetic topological Weyl semimetals (WSM) [25-27] as well as in non-collinear AFM systems with strong spin-orbit coupling and broken time-reversal and lattice symmetries [28].

Since Weyl nodes are singular points of Berry curvature, a large net value of the Berry curvature can occur in time-reversal-symmetry (TRS) breaking WSMs, which can result in strong AHE when the Weyl nodes are close to the Fermi level [26]. This scenario has been demonstrated recently in ferromagnetic (FM) WSMs $Co_3Sn_2S_2$ [29-32] and $Co_2MnGa$ [33-35], where the anomalous Hall angles (AHA) reach values of 0.2 and 0.1, respectively, which are far larger than those observed in other magnetic conductors. WSM states can also be present in AFM materials such as GdPtBi [36,37], TbPtBi [38], $Mn_3Sn$ and $Mn_3Ge$ [39-43]. In particular, GdPtBi and TbPtBi show a large AHE with AHA values of 0.16-0.38 [37,38,44]. Previous studies of GdPtBi suggest two possible mechanisms for its large AHE. One explanation is that the spin texture under magnetic field leads to an avoided band-crossing near the Fermi energy $E_f$, which can drive a significant Berry curvature contribution to the AHE [44-46], while another explanation involves the contribution to the Berry curvature from field-induced Weyl points [37,38,44]. However, it remains unclear whether the Weyl node or the avoided band-crossing is the key contribution for inducing the large AHE. In this article, we show that an exceptionally large intrinsic AHE with anomalous Hall angles of 0.68-0.76 can be created via the anti-crossing of inverted, spin-split bands in TbPtBi, and that it can be probed when the magnetic field is applied along the [100] direction. We further show that Weyl points hardly contribute to the AHE under this field orientation.

Like GdPtBi, TbPtBi belongs to the half-Heusler family. We have grown TbPtBi single crystals using Bi-flux method, see Supplementary Materials (SM) for details [47]. Our TbPtBi single crystals exhibit semi-metallic behavior with a broad peak in resistivity $\rho_{xx}$ around 225K (Fig. 1a). The magnetic susceptibility ($\chi$) measurements (Fig. 1b) reveal an AFM transition with a Neel temperature of $T_N$ = 3.4 K. The $\chi(T)$ in the high-temperature range (60-300K) can be fitted by the Curie-Weiss form; the effective magnetic moment ($\mu_{eff}$) extracted from the fit is ~ 9.41 $\mu_B$/Tb, see Supplementary Fig. S1. The $T_N$ and $\mu_{eff}$ values are consistent with an earlier report [38].

In the Hall effect measurements, we observed an intrinsic AHE, much stronger than that seen in previous studies on TbPtBi [38] and GdPtBi [37,44]. Our measurements were conducted with the magnetic field applied along the [100] axis (inset, Fig. 1a) and the current along the [010] or [001] direction ([010] and [001] are equivalent in a cubic structure), while previous experiments on TbPtBi and GdPtBi involved $I$//[110] and $B$//[111]. Figure 1c shows the Hall resistivity ($\rho_{xy}$) of a typical TbPtBi sample as a function of the magnetic field at various temperatures. Below 20K, a clear anomalous feature emerges, which is analogous to what has been seen in GdPtBi [37,44]. $\rho_{xy}$ deviates from linear field dependence and the magnetic field $H^*$ where $\rho_{xy}$ deviates from linearity decrease with decreasing temperature, ~ down to 3-4T at 2K. We also find that the anomalous Hall resistivity is not linearly coupled to the magnetization *M*. The *M-H* curve shows a slightly deviated linear behavior in the AFM state at 2K whereas it is linear in the paramagnetic (PM) state at 9K as shown in the inset of Fig. 1c. We further examine the magnetization dependence of anomalous Hall resistivity $\Delta\rho_{xy}$ which is obtained by subtracting normal Hall contribution $R_0B$ ($R_0$, Hall coefficient) and shown in the inset of Fig. 1d. It can be seen clearly that $\Delta\rho_{xy}$ vs. *M* does not follows a linear *M* dependence as expected in trivial magnetic conductors, but shows a significant peak near the field where $\rho_{xy}$ deviates from linear field dependence, suggesting that the large AHE observed in TbPtBi is of intrinsic origin.

To estimate the intrinsic anomalous Hall conductivity $\sigma_{yx}^A$, we followed the approach of Ref. [44] and define $\sigma_{yx}^A = \sigma_{yx} - \sigma_{yx}^N$, where $\sigma_{yx}^N$ is the normal Hall conductivity, which can be estimated using the Hall resistivity $\rho_{xy}$ and longitudinal resistivity $\rho_{xx}$ measured at a temperature far above $T_N$ (e.g. 20K). Both $\sigma_{yx}$ and $\sigma_{yx}^N$ are obtained through tensor conversions from the measured $\rho_{xx}$ and $\rho_{xy}$, i.e. $\sigma_{yx} = \rho_{xy}/(\rho_{xx}^2+\rho_{xy}^2)$. The calculated $\sigma_{yx}^A$ is presented in the Supplementary Fig. S2a where we can see that the maximal value of $\sigma_{yx}^A$ reaches ~ 1,100 $\Omega^{-1}$cm$^{-1}$ near 1T/4T at 2K, much larger than the previously reported values for GdPtBi (30-200 $\Omega^{-1}$cm$^{-1}$ [44]) and TbPtBi (~744 $\Omega^{-1}$cm$^{-1}$ [38]). In Fig. 1d, we present the anomalous Hall angle ($\theta_{AH} = \sigma_{yx}^A/\sigma_{xx}$) as a function of the magnetic field. We find the maximal $\theta_{AH}$ value of this sample to be as large as ~0.68 near 4T at 2K, about 1.8 times of the previously reported $\theta_{AH}$ (=0.38 [38]) for TbPtBi and ~2-5 times larger than the values seen in GdPtBi ($\theta_{AH}$ = 0.16-0.32 [44] ). The Hall angle of $\theta_{AH}$ = 0.38 for TbPtBi was claimed to the largest among other materials [38]; our result thus sets a new record. Such a large $\theta_{AH}$ is verified in other samples. In Supplementary Fig. S2b, we show the Hall angle measured on another sample whose maximal $\theta_{AH}$ is ~ 0.76, even larger than that of the sample shown in Fig. 1d.

Although we take the Hall conductivity at 20K as $\sigma_{yx}^N$, we note that the field dependence of $\rho_{xy}$ is not exactly linear at 20K but becomes almost linear at 50K. The reason why we did not use the Hall conductivity data at 50K for normal Hall contribution subtraction is that $R_0$ in the PM state is weakly temperature dependent. Nevertheless, we find $\rho_{xy}$ collapses to a linear field dependence above 7.5T for $T \leq 20K$, implying the variation of $R_0$ is very small below 20K. The non-linear field dependence of $\rho_{xy}$ below 7.5T at 20K implies weak AHE due to spin fluctuations. Thus, using the Hall conductivity of 20K for normal Hall subtraction may lead to small underestimates of $\sigma_{yx}^A$ and $\theta_{AH}$. We also attempted using the Hall conductivity data of 50K for normal Hall subtraction and found the $\sigma_{yx}^A$ and $\theta_{AH}$ values are 1347 $\Omega^{-1}cm^{-1}$ and 0.86, respectively, at 2K (see supplementary Fig. S4), greater than those values obtained using the Hall conductivity of 20K for normal Hall subtraction. Of course, the temperature dependence of $R_0$ may cause small overestimates of $\sigma_{yx}^A$ and $\theta_{AH}$ in this case.

The previously reported large $\theta_{AH}$ for TbPtBi was attributed to the field-induced Weyl state [38]. The main support for this argument is the observation of the negative longitudinal magnetoresistance (LMR), which was interpreted as arising from the chiral anomaly of a field-induced Weyl state. Negative LMR was also observed in our experiments with $H//[100]$, as shown in Fig. 2a. Our LMR data also display a dip feature near 4T for $T<T_N$, which becomes more significant when the field tilt angle is above 30°. Although all these features look similar to those seen in GdPtBi [36,37] and appear to be consistent with a field-induced Weyl state, we observed signatures unexpected for a Weyl state in planar Hall effect (PHE) measurements. PHE refers to the Hall effect which takes place when the magnetic field and the electrical current are co-planar (inset, Fig. 2c). Such a Hall effect can be caused by the magnetoresistivity anisotropy that usually occurs in ferromagnets due to anisotropic magnetization. In Weyl semimetals, the chiral anomaly leads to a large magnetoresistivity anisotropy, i.e. the resistivity measured with $B // I$ ($\rho_{//}$) differs from the resistivity probed with $B \perp I$ ($\rho_{\perp}$). The Hall resistivity of PHE ($\rho_{xy}^{PHE}$) induced by the chiral anomaly can be expressed as

$$\rho_{xy} = \frac{(\rho_{\parallel} - \rho_{\perp})}{2} \sin 2\varphi \quad (1),$$

where $\varphi$ is the angle between the current flow and magnetic field orientation (inset, Fig. 2c) and $\rho_{//}-\rho_{\perp}$ represents the resistivity anisotropy caused by the chiral anomaly [48]. PHE has been used as an alternative probe for the chiral anomaly, and it has been observed in a number of topological semimetals such as $ZrTe_5$, $Cd_3As_2$, GdPtBi, $WTe_2$, $MoTe_2$ and $VAl_3$ [49-54]. As shown in Fig. 2c, the $\rho_{xy}^{PHE}$ measured in our TbPtBi sample shows a very complex field-dependent anisotropic behavior. We did not observe the expected $\sin 2\varphi$ angular dependence in PHE due to chiral anomaly at low fields, but a complex anisotropy with minima occurring at 45° and 225° (e.g. see the data taken at 4T in Fig. 2c). However, at higher fields (e.g. 6T and 8T), a $\sin 2\varphi$ component appears to be superimposed on the anisotropic component seen at 4T. This suggests that the PHE revealed in our experiments may involve both the chiral anomaly and anisotropic magnetic scattering effect. The anisotropic magnetic scattering is also clearly manifested in the in-plane angular dependence of $\rho_{xx}$ where it exhibits oscillations with field-dependent symmetry in the AFM state (Fig. 2d). At 4-6T, it shows 2/4-fold oscillations with the maxima occurring at $B//I$ (i.e. $\phi = 0$ and 180°) in contrast to the expected chiral-anomaly induced minimal in $\rho_{xx}$ for $B//I$ [36] and $(\cos\phi)^2$ dependence of $\rho_{xx}(\phi)$ [48]. When the field is increased to 8T, $\rho_{xx}(\phi)$ shows more complex oscillations with 8-fold symmetry and local minima at $B//I$, implying the possible superposition of chiral anomaly and magnetic scattering. We have also measured $\rho_{xx}(\phi)$ in the PM state at 10K (the top curve in Fig. 2d) and find it exhibits minima for both $B // I$ and $B \perp I$ that is inconsistent with the chiral anomaly effect. These observations suggest that the negative LMR shown in Fig. 2a mostly arise from magnetic scattering with the possibility of a relatively small contribution from the chiral anomaly.

In order to understand the AHE of TbPtBi, we have performed detailed band structure calculations, see SM for the details. According to previous electronic structure calculations on half-Heusler materials [55-57], the combination of Pt with Bi in REPtBi (RE = rare earth) leads to an inverted band structure where the $\Gamma_6$ and $\Gamma_8$ symmetry bands are inverted at Γ. Since the band-inversion-strength (BIS) increases with the average nuclear charge of the atoms in the unit cell, we expect TbPtBi to have a greater BIS value than GdPtBi. Our first-principles band structure calculations indeed show that TbPtBi and GdPtBi share a similar band structure and that the BIS value is larger in TbPtBi at zero magnetic field.

The arrangement of atoms in the unit cell in TbPtBi is illustrated in Figs. 3a and 3b. For the nonmagnetic state, it is described by the inversion asymmetric space group $T_d^2$ ($F\bar{4}3m$, $No.$ 216) where Tb, Pt, and Bi atoms occupy Wykoff positions 4b, 4c, and 4a, respectively. The crystal structure consists of a layered atomic arrangement when viewed along [111] with $Tb^{3+}$ ions stuffing the $[PtBi]^{3-}$ zinc-blende sublattice at the remaining sites of a face-centered cubic (fcc) lattice. Notably, the REPtBi (RE=rare earth) compounds generally possess two different types of AFM orders. The materials with lighter RE atoms such as CePtBi and NdPtBi tend to from type-I AFM order with the propagation vector along [100] [57-58]. In contrast, the materials with heavier RE atoms such as GdPtBi and YbPtBi form a type-II AFM order with the propagation vector [111] [44, [58]. Since Tb is heavier than Gd and GdPtBi exhibits type-II AFM order, we consider a type-II AFM order for TbPtBi in our calculations. For the type-II AFM ordering, the space group symmetry is reduced to $C_{3v}^5 (R3m,\ No.\ 160)$ as seen in GdPtBi. This crystal structure is viewed as trigonal in which Tb magnetic moments align ferromagnetically in an atomic plane perpendicular to [111] and antiferromagnetically between the two neighboring planes (see Fig. 3b). This structure breaks inversion symmetry but preserves an effective time-reversal symmetry $S = \Theta \tau_{1/2}$, where $\Theta$ is the time-reversal operator and $\tau_{1/2}$ is the half-translation vector connecting spin-up and spin-down Tb atoms. We find that Tb ion has a net effective magnetic moment of 9.31 $\mu_B$/Tb which is composed of a spin magnetic moment of 6.07 $\mu_B$ and orbital moment of 3.24 $\mu_B$, consistent with a simple $Tb^{3+}$ ionic state and the experimental value of $\mu_{\text{eff}} \sim 9.41\ \mu_B$/Tb.

The bulk band structure of TbPtBi in the nonmagnetic state is semimetallic, exhibiting a $\Gamma_8 (p-\text{type})$ and $\Gamma_6 (s-\text{type})$ band inversion similar to other half-Heusler materials such as GdPtBi. On considering the type-II AFM spin-order, TbPtBi retains the band-inverted semimetallic features of the nonmagnetic state as shown in Fig. 3d without the SOC. When we include SOC, the nodal crossings vanish, with a bandgap opening at the band-crossing points, see Fig. 3e. A careful inspection of the band structure in the full BZ reveals that the system has a continuous gap between the valence and conduction bands. However, the band structure remains semimetallic with two large spin-split hole-like Fermi pockets around Γ as shown in Fig. 3c. The Fermi wave vector ($k_F$) for the inner (outer) pocket along the Γ-L, Γ-F, and Γ-Z lines is ~ 0.053 (0.066) Å$^{-1}$, 0.083 (0.125) Å$^{-1}$, and 0.091 (0.135) Å$^{-1}$, respectively, comparable to $k_F$ probed in Shubnikov-de Haas quantum oscillation experiments, see Supplementary Fig. S3. Notably, quantum oscillations probe a single anisotropic hole pocket. This is because the large exchange coupling in the presence of an external magnetic field pushes the second pocket below the Fermi level as seen in Fig. 4. Although TbPtBi is a semimetal, owing to the existence of the SOC-induced bandgap locally at each *k* point and an effective time-reversal symmetry *S*, we can define a *Z₂* topological invariant in a manner similar to the time-reversal symmetric topological insulators. By tracking the evolution of the Wannier charge centers on the $k_i = 0$ and π planes, we obtained *Z₂* topological invariant to be 1, concluding that TbPtBi is an antiferromagnetic topological insulator. Notably, this is in strong contrast to GdPtBi which remains in a nontrivial semimetal state even after considering SOC in the computations.

We turn next to consider the ferromagnetic structure of TbPtBi and reveal the emergence of the Weyl semimetal state for *B*||[100] in Fig. 4. When Tb magnetic moments are aligned ferromagnetically, the spin-degenerate bands split in energy due to the strong exchange coupling. By calculating the splitting between the majority and minority spin states at *Γ*, we estimate the exchange coupling to be ~0.4 eV. As shown in Fig. 4a, the valence and conduction states of the different spin-channels cross near the Fermi level, forming a nodal surface. When we include SOC with a [100] magnetization, the symmetry of TbPtBi lattice reduces to *S₄*. We find that the spin-degeneracy at the nodal band-crossings is lifted everywhere in the BZ except at the four *k* points which are listed in Table S1 in SM. These points constitute two pairs of *S₄* symmetry-related $W_{100}^1$ type-II Weyl nodes and are located at 83 meV above the *E_f* as illustrated in Fig. 4c, contrasted with 6 pairs of Weyl nodes in GdPtBi under the same field orientation. The Weyl node separation is ~0.16 Å⁻¹, much smaller than that in GdPtBi [36,37]. These results clearly show that TbPtBi supports a magnetic field induced topological phase transition from a nontrivial AFM insulator to a Weyl semimetal. Given that the calculated Weyl nodes are well above *E_f* and that the anomalous Hall conductivity (AHC) is in proportion to the Weyl node separation Δ*k*_w [26,59], it follows that the predicted Weyl points do not make a significant contribution to the large AHE observed in TbPtBi, consistent with our experimental results which also do not reveal remarkable transport signatures of a field-induced Weyl state as discussed above.

In order to gain further insight into the nature of the unusually large anomalous Hall response of TbPtBi, we have calculated the Berry curvature distribution associated with our band structure. Fig. 4d presents the Berry curvature distribution along the selected high-symmetry directions. The Berry curvature around the Fermi level is seen to be predominantly positive for valence bands (hole pockets) and negative for the conduction bands (electron pockets). At $E_f$, the Berry curvature contains both positive and negative hot-spots as shown in Fig. 4e. However, the size of negative hot-spots is much smaller compared to the positive hot-spots. Through the Berry curvature integrations, such a distribution of opposite hot-spots is found to produce a large AHC that can be seen in our calculated AHC $\sigma_{yx}^A$ (Fig. 4f). The $\sigma_{yx}^A$ maintains a plateau from $E_f$ to -45 meV with a maximum value of ~ 320 $\Omega^{-1}$cm$^{-1}$ showing predominantly positive Berry curvature. On increasing the energy above $E_f$, $\sigma_{yx}^A$ drops quickly and changes sign at ~ 140 meV. Although the calculated $\sigma_{yx}^A$ near $E_f$ is less than the maximal experimental value (1100 $\Omega^{-1}$cm$^{-1}$), it clearly indicates that Berry curvature from the nontrivial hole-like Fermi pockets plays a dominant role in the observed large AHC. The formation of such a hole pocket is driven by the anti-crossing of spin-split bands, i.e. gapping of the nodal surface. The presence of strong SOC and the lattice symmetry reduction caused by the magnetic order play a key role in triggering such band anti-crossing and creating large net Berry curvature.

The difference between the calculated and experimental values of AHC is due to the fact that our calculations are based on a fully-polarized FM state, while the actual magnetic state near ~4T where the maximal $\sigma_{yx}^A$ is observed shows a canted AFM order. This implies that the non-collinear spin structure enhances the Berry curvature, which is consistent with an earlier theoretical prediction that a large net Berry curvature can be created by the symmetry breaking plus strong SOC in a magnetic state with non-collinear spin structure [28]. As noted above, the large AHE observed in GdPtBi was attributed to avoided band crossings or Weyl points [44]. Given the similarity of band structures and the differences in the Weyl states between GdPtBi and TbPtBi as discussed above, the band anti-crossing may also play a key role in creating large AHE in GdPtBi.

In summary, we have observed a strong intrinsic AHE effect with an exceptionally large anomalous Hall angle (0.68-0.76) in TbPtBi, which cannot be attributed to field-induced Weyl points. Through Berry curvature and AHC calculations, we find that the non-trivial hole pocket formed by band anti-crossing carries large Berry curvature and drives the large AHE in TbPtBi. The theoretical analysis also suggests that TbPtBi is an AFM topological insulator in its ground state. Our study thus advances the understanding of exotic quantum transport properties induced by non-trivial band topologies in materials.

**Acknowledgment**:

This work is supported by the US National Science Foundation under grants DMR 1917579 and 1832031. A portion of this work was performed at the National High Magnetic Field Laboratory, which is supported by National Science Foundation Cooperative Agreement No. DMR-1157490 and the State of Florida. Y.L.Z and Z.Q.M. acknowledge financial support from the National Science Foundation through the Penn State 2D Crystal Consortium-Materials Innovation Platform (2DCC-MIP) under NSF cooperative agreement DMR-1539916.The work at Northeastern University was supported by the US Department of Energy (DOE), Office of Science, Basic Energy Sciences Grant No. DE-FG02-07ER46352, and benefited from Northeastern University's Advanced Scientific Computation Center and the National Energy Research Scientific Computing

Center through DOE Grant No. DE-AC02-05CH11231. H.L. acknowledges Academia Sinica (Taiwan) for support under Innovative Materials and Analysis Technology Exploration (AS-iMATE-107-11).

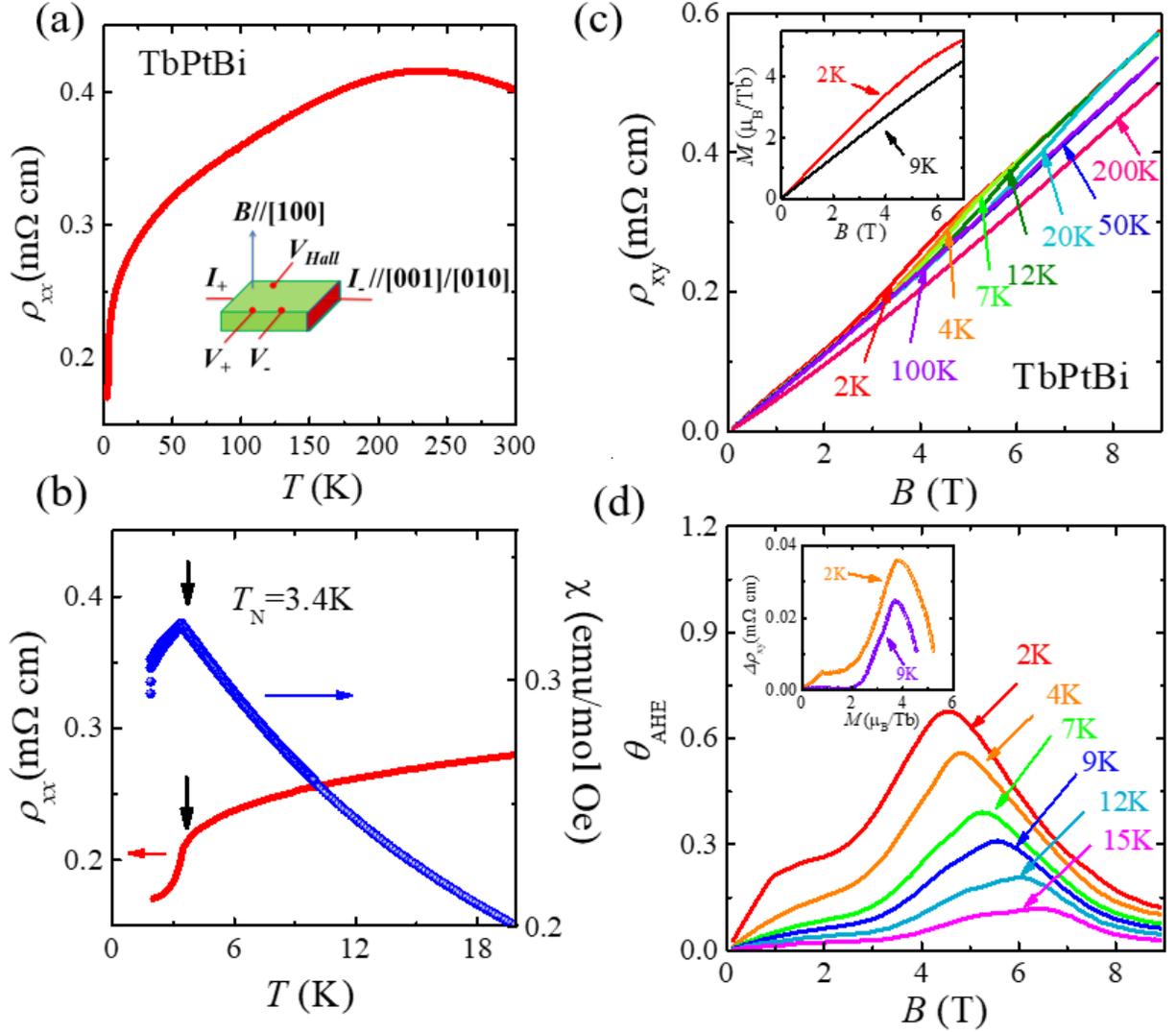

**Figure 1**. (a) Temperature dependence of resistivity of TbPtBi. Inset, the schematic of experimental set up for longitudinal and Hall resistivity measurements. (b) The AFM transition probed in the resistivity and the magnetic susceptibility measured with $B = 0.1T(B//[100])$. (c) Hall resistivity as a function of the magnetic field $B$ for TbPtBi. The inset in (c) shows the isothermal magnetization data of TbPtBi at $T$=2K and 9K. (d) Anomalous Hall angle vs. field at various temperatures. The inset in (d) shows the magnetization dependence of the anomalous Hall resistivity $\Delta\rho_{xy}$ (which is obtained by subtracting normal Hall contribution $R_0B$) at 2K and 9K.

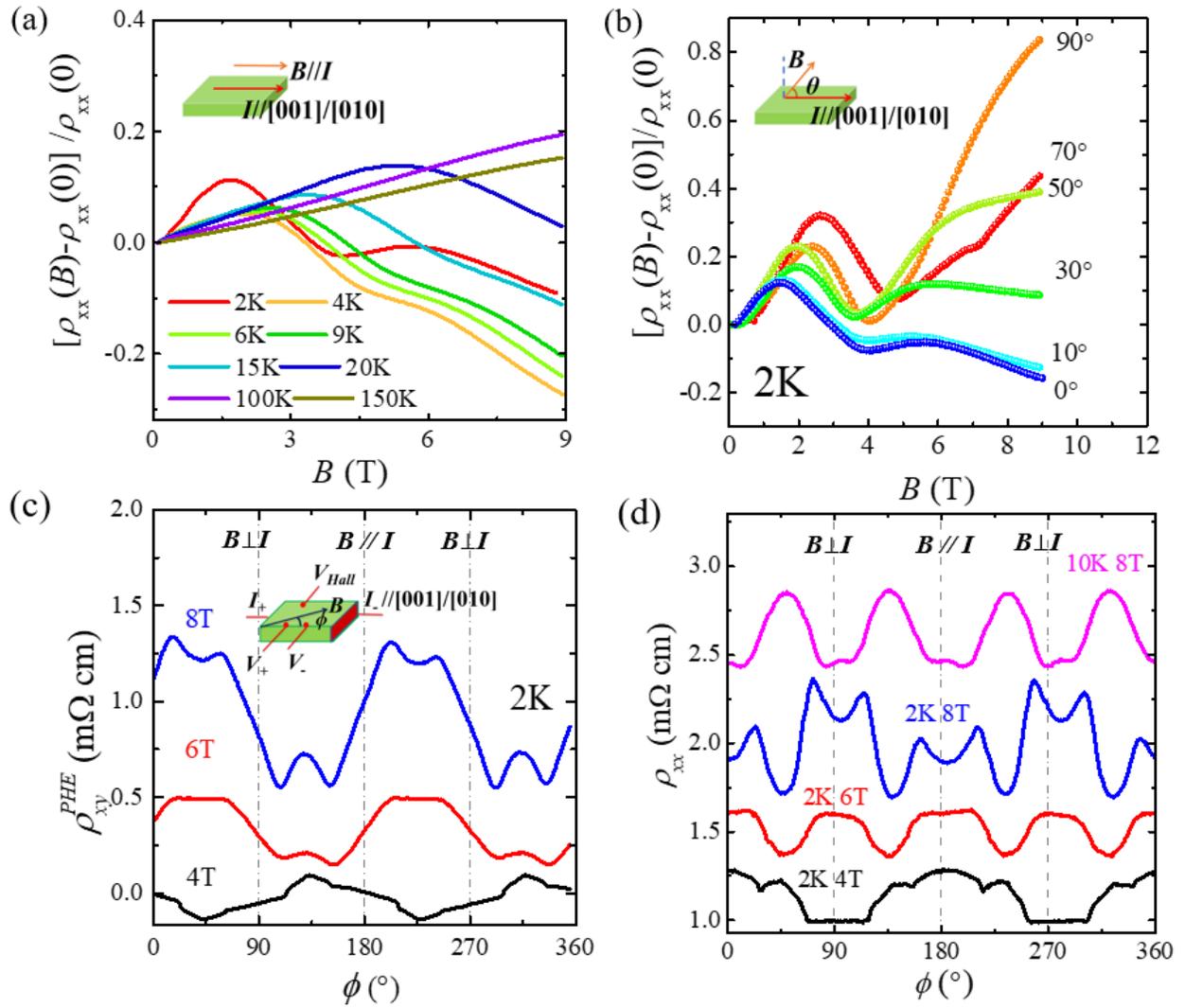

**Figure 2**. (a) Field dependence of longitudinal magnetoresistivity at various temperatures. (b) Field dependence of magnetoresistivity at 2K under various field orientations. (c) Angular dependence of planar Hall resistivity $\rho_{xy}^{PHE}$ under various magnetic fields at $T = 2K$. The data for 6 and 8T have been shifted for clarity. The insets in panels (a), (b) and (c) show schematics of experimental setups. (d) Angular dependence of $\rho_{xx}$ under different magnetic fields at 2K. The data for 6T and 8T at 2/10K have been shifted for clarity.

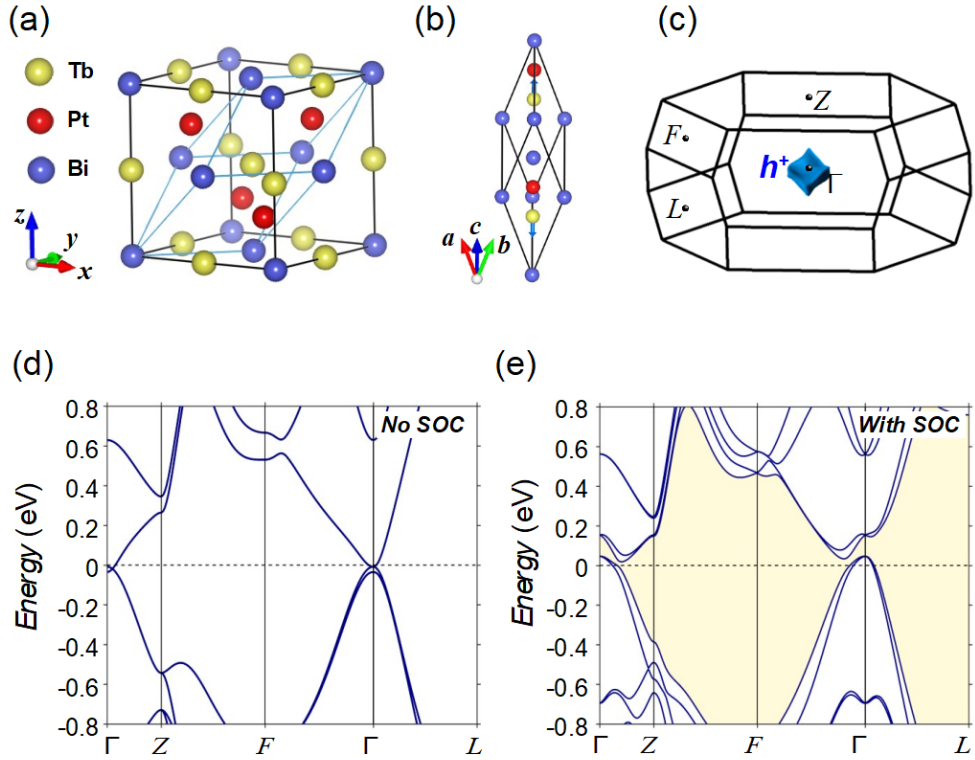

**Figure 3.** (a) Crystal structure of the half-Heusler TbPtBi. The conventional and primitive unit cells are denoted by black and blue lines. (b) The rhombohedral unit cell for the type-II antiferromagnetically ordered TbPtBi. (c) Fermi surface of the type-II antiferromagnetically ordered rhombohedral TbPtBi, calculated with the inclusion of spin-orbit coupling (SOC). Hole-like Fermi surface at $\Gamma$ is colored in blue. Bulk band structure of the type-II AFM TbPtBi (d) without and (e) with SOC. The shaded yellow region in (e) highlights the continuous gap between the valence and conduction states.

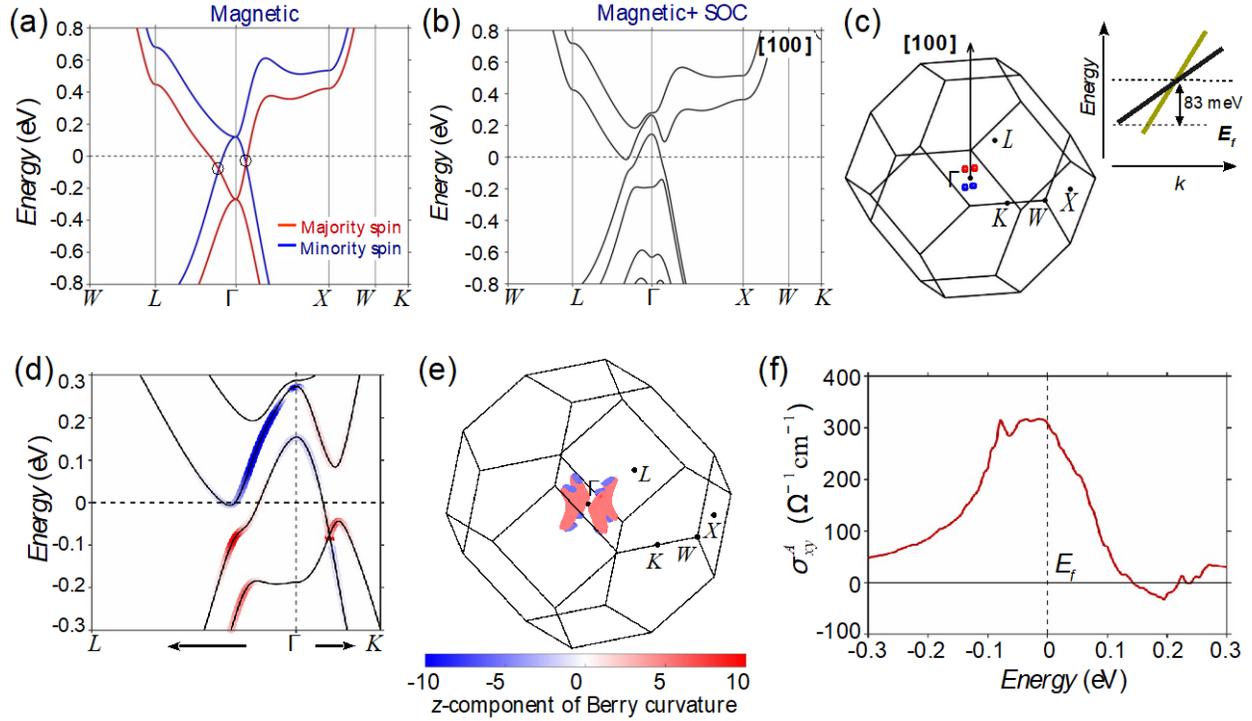

**Figure 4.** (a) Spin-resolved band structure of ferromagnetically ordered half-Heusler TbPtBi without SOC. Crossings between the valence and conduction bands from different spin-channels are highlighted with black circles. (b) Band structure with the inclusion of SOC where the magnetic moments are fully aligned along the [100] direction. (c) Distribution of the four Weyl nodes in the first BZ of TbPtBi. Red and blue spheres represent the Weyl nodes with '+1' and '-1' chiral charge, respectively. The right inset is a schematic of the band dispersion of Weyl fermions. (d)-(e) Berry curvature distribution for TbPtBi along the selected high-symmetry lines and in the full BZ at the Fermi level. (f) Energy dependence of the anomalous Hall conductivity of TbPtBi with all moments oriented along the [100] direction. The vertical broken line marks the Fermi energy.

References:

[1] R. Karplus and J. M. Luttinger, Hall effect in ferromagnetics. Phys. Rev. **95**, 1154 (1954).
[2] J. Smit, The spontaneous hall effect in ferromagnetics II. Physica **24**, 39 (1958).
[3] L. Berger, Side-Jump Mechanism for the Hall Effect of Ferromagnets. Phys. Rev. B **2**, 4559 (1970).
[4] N. Nagaosa, J. Sinova, S. Onoda, A. H. MacDonald, and N. P. Ong, Anomalous Hall effect. Rev. Mod. Phys. **82**, 1539 (2010).
[5] J. Smit, The spontaneous hall effect in ferromagnetics I. Physica **21**, 877 (1955).
[6] P. Matl, N. P. Ong, Y. F. Yan, Y. Q. Li, D. Studebaker, T. Baum, and G. Doubinina, Hall effect of the colossal magnetoresistance manganite $La_{1-x}Ca_xMnO_3$. Phys. Rev. B **57**, 10248 (1998).
[7] K. Ohgushi, S. Murakami, and N. Nagaosa, Spin anisotropy and quantum Hall effect in the kagome lattice: Chiral spin state based on a ferromagnet. Phys. Rev. B **62**, R6065 (2000).
[8] J. Ye, Y. B. Kim, A. J. Millis, B. I. Shraiman, P. Majumdar, and Z. Tešanović, Berry Phase Theory of the Anomalous Hall Effect: Application to Colossal Magnetoresistance Manganites. Phys. Rev. Lett **83**, 3737 (1999).
[9] Y. Taguchi, Y. Oohara, H. Yoshizawa, N. Nagaosa, and Y. Tokura, Spin Chirality, Berry Phase, and Anomalous Hall Effect in a Frustrated Ferromagnet. Science **291**, 2573 (2001).
[10] Y. Machida, S. Nakatsuji, Y. Maeno, T. Tayama, T. Sakakibara, and S. Onoda, Unconventional Anomalous Hall Effect Enhanced by a Noncoplanar Spin Texture in the Frustrated Kondo Lattice $Pr_2Ir_2O_7$. Phys. Rev. Lett **98**, 057203 (2007).
[11] W.-L. Lee, S. Watauchi, V. L. Miller, R. J. Cava, and N. P. Ong, Dissipationless anomalous Hall current in the ferromagnetic spinel $CuCr_2Se_{4-x}Br_x$. Science **303**, 1647 (2004).
[12] N. Manyala, Y. Sidis, J. F. DiTusa, G. Aeppli, D. P. Young, and Z. Fisk, Large anomalous Hall effect in a silicon-based magnetic semiconductor. Nat. Mat. **3**, 255 (2004).
[13] M. Lee, Y. Onose, Y. Tokura, and N. P. Ong, Hidden constant in the anomalous Hall effect of high-purity magnet MnSi. Phys. Rev. B **75**, 172403 (2007).
[14] A. Neubauer, C. Pfleiderer, B. Binz, A. Rosch, R. Ritz, P. G. Niklowitz, and P. Böni, Topological Hall Effect in the A Phase of MnSi. Phys. Rev. Lett **102**, 186602 (2009).
[15] F. Freimuth, R. Bamler, Y. Mokrousov, and A. Rosch, Phase-space Berry phases in chiral magnets: Dzyaloshinskii-Moriya interaction and the charge of skyrmions. Phys. Rev. B **88**, 214409 (2013).
[16] F. Freimuth, S. Blügel, and Y. Mokrousov, Berry phase theory of Dzyaloshinskii–Moriya interaction and spin–orbit torques. J. Phys.: Condens. Matter **26**, 104202 (2014).
[17] C. Franz, F. Freimuth, A. Bauer, R. Ritz, C. Schnarr, C. Duvinage, T. Adams, S. Blügel, A. Rosch, Y. Mokrousov, and C. Pfleiderer, Real-Space and Reciprocal-Space Berry Phases in the Hall Effect of $Mn_{1-x}Fe_xSi$. Phys. Rev. Lett **112**, 186601 (2014).
[18] T. Jungwirth, Q. Niu, and A. H. MacDonald, Anomalous Hall effect in ferromagnetic semiconductors. Phys. Rev. Lett **88**, 207208 (2002).
[19] M. Onoda and N. Nagaosa, Topological Nature of Anomalous Hall Effect in Ferromagnets. J. Phys. Soc. Jpn. **71**, 19 (2002).
[20] M. Onoda and N. Nagaosa, Quantized anomalous Hall effect in two-dimensional ferromagnets: quantum Hall effect in metals. Phys. Rev. Lett **90**, 206601 (2003).17

[21]   Z. Fang, N. Nagaosa, K. S. Takahashi, A. Asamitsu, R. Mathieu, T. Ogasawara, H. Yamada, M. Kawasaki, Y. Tokura, and K. Terakura, The Anomalous Hall Effect and Magnetic Monopoles in Momentum Space. Science **302**, 92 (2003).

[22]   F. D. M. Haldane, Berry Curvature on the Fermi Surface: Anomalous Hall Effect as a Topological Fermi-Liquid Property. Phys. Rev. Lett **93**, 206602 (2004).

[23]   Y. Yao, L. Kleinman, A. H. MacDonald, J. Sinova, T. Jungwirth, D.-s. Wang, E. Wang, and Q. Niu, First Principles Calculation of Anomalous Hall Conductivity in Ferromagnetic bcc Fe. Phys. Rev. Lett **92**, 037204 (2004).

[24]   D. Xiao, J. Shi, and Q. Niu, Berry Phase Correction to Electron Density of States in Solids. Phys. Rev. Lett **95**, 137204 (2005).

[25]   J. F. Steiner, A. V. Andreev, and D. A. Pesin, Anomalous Hall Effect in Type-I Weyl Metals. Phys. Rev. Lett **119**, 036601 (2017).

[26]   A. A. Burkov, Anomalous Hall Effect in Weyl Metals. Phys. Rev. Lett **113**, 187202 (2014).

[27]   A. A. Zyuzin and R. P. Tiwari, Intrinsic anomalous Hall effect in type-II Weyl semimetals. JETP Letters **103**, 717 (2016).

[28]   H. Chen, Q. Niu, and A. H. MacDonald, Anomalous Hall Effect Arising from Noncollinear Antiferromagnetism. Phys. Rev. Lett **112**, 017205 (2014).

[29]   E. Liu, Y. Sun, N. Kumar, L. Muechler, A. Sun, L. Jiao, S.-Y. Yang, D. Liu, A. Liang, Q. Xu, J. Kroder, V. Süß, H. Borrmann, C. Shekhar, Z. Wang, C. Xi, W. Wang, W. Schnelle, S. Wirth, Y. Chen, S. T. B. Goennenwein, and C. Felser, Giant anomalous Hall effect in a ferromagnetic kagome-lattice semimetal. Nature Physics **14**, 1125 (2018).

[30]   Q. Wang, Y. Xu, R. Lou, Z. Liu, M. Li, Y. Huang, D. Shen, H. Weng, S. Wang, and H. Lei, Large intrinsic anomalous Hall effect in half-metallic ferromagnet $Co_3Sn_2S_2$ with magnetic Weyl fermions. Nat. Commun. **9**, 3681 (2018).

[31]   D. F. Liu, A. J. Liang, E. K. Liu, Q. N. Xu, Y. W. Li, C. Chen, D. Pei, W. J. Shi, S. K. Mo, P. Dudin, T. Kim, C. Cacho, G. Li, Y. Sun, L. X. Yang, Z. K. Liu, S. S. P. Parkin, C. Felser, and Y. L. Chen, Magnetic Weyl semimetal phase in a Kagomé crystal. Science **365**, 1282 (2019).

[32]   N. Morali, R. Batabyal, P. K. Nag, E. Liu, Q. Xu, Y. Sun, B. Yan, C. Felser, N. Avraham, and H. Beidenkopf, Fermi-arc diversity on surface terminations of the magnetic Weyl semimetal $Co_3Sn_2S_2$. Science **365**, 1286 (2019).

[33]   A. Sakai, Y. P. Mizuta, A. A. Nugroho, R. Sihombing, T. Koretsune, M.-T. Suzuki, N. Takemori, R. Ishii, D. Nishio-Hamane, R. Arita, P. Goswami, and S. Nakatsuji, Giant anomalous Nernst effect and quantum-critical scaling in a ferromagnetic semimetal. Nature Physics **14**, 1119 (2018).

[34]   S. N. Guin, K. Manna, J. Noky, S. J. Watzman, C. Fu, N. Kumar, W. Schnelle, C. Shekhar, Y. Sun, J. Gooth, and C. Felser, Anomalous Nernst effect beyond the magnetization scaling relation in the ferromagnetic Heusler compound $Co_2MnGa$. NPG Asia Materials **11**, 16 (2019).

[35]   I. Belopolski, K. Manna, D. S. Sanchez, G. Chang, B. Ernst, J. Yin, S. S. Zhang, T. Cochran, N. Shumiya, H. Zheng, B. Singh, G. Bian, D. Multer, M. Litskevich, X. Zhou, S.-M. Huang, B. Wang, T.-R. Chang, S.-Y. Xu, A. Bansil, C. Felser, H. Lin, and M. Z. Hasan, Discovery of topological Weyl fermion lines and drumhead surface states in a room temperature magnet. Science **365**, 1278 (2019).

[36]     M. Hirschberger, S. Kushwaha, Z. Wang, Q. Gibson, S. Liang, C. A. Belvin, B. A. Bernevig, R. J. Cava, and N. P. Ong, The chiral anomaly and thermopower of Weyl fermions in the half-Heusler GdPtBi. Nat. Mat. **15**, 1161 (2016).
[37]     C. Shekhar, N. Kumar, V. Grinenko, S. Singh, R. Sarkar, H. Luetkens, S.-C. Wu, Y. Zhang, A. C. Komarek, E. Kampert, Y. Skourski, J. Wosnitza, W. Schnelle, A. McCollam, U. Zeitler, J. Kübler, B. Yan, H. H. Klauss, S. S. P. Parkin, and C. Felser, Anomalous Hall effect in Weyl semimetal half-Heusler compounds RPtBi (R = Gd and Nd). Proc. Natl. Acad. Sci. USA **115**, 9140 (2018).
[38]     R. Singha, S. Roy, A. Pariari, B. Satpati, and P. Mandal, Magnetotransport properties and giant anomalous Hall angle in the half-Heusler compound TbPtBi. Phys. Rev. B **99**, 035110 (2019).
[39]     S. Nakatsuji, N. Kiyohara, and T. Higo, Large anomalous Hall effect in a non-collinear antiferromagnet at room temperature. Nature **527**, 212 (2015).
[40]     A. K. Nayak, J. E. Fischer, Y. Sun, B. Yan, J. Karel, A. C. Komarek, C. Shekhar, N. Kumar, W. Schnelle, J. Kübler, C. Felser, and S. S. P. Parkin, Large anomalous Hall effect driven by a nonvanishing Berry curvature in the noncolinear antiferromagnet $Mn_3Ge$. Science Advances **2**, e1501870 (2016).
[41]     J. Kübler and C. Felser, Weyl fermions in antiferromagnetic $Mn_3Sn$ and $Mn_3Ge$. Europhys. Lett. **120**, 47002 (2017).
[42]     H. Yang, Y. Sun, Y. Zhang, W.-J. Shi, S. S. P. Parkin, and B. Yan, Topological Weyl semimetals in the chiral antiferromagnetic materials $Mn_3Ge$ and $Mn_3Sn$. New J. Phys. **19**, 015008 (2017).
[43]     K. Kuroda, T. Tomita, M. T. Suzuki, C. Bareille, A. A. Nugroho, P. Goswami, M. Ochi, M. Ikhlas, M. Nakayama, S. Akebi, R. Noguchi, R. Ishii, N. Inami, K. Ono, H. Kumigashira, A. Varykhalov, T. Muro, T. Koretsune, R. Arita, S. Shin, T. Kondo, and S. Nakatsuji, Evidence for magnetic Weyl fermions in a correlated metal. Nat. Mat. **16**, 1090 (2017).
[44]     T. Suzuki, R. Chisnell, A. Devarakonda, Y. T. Liu, W. Feng, D. Xiao, J. W. Lynn, and J. G. Checkelsky, Large anomalous Hall effect in a half-Heusler antiferromagnet. Nature Physics **12**, 1119 (2016).
[45]     K. Manna, L. Muechler, T.-H. Kao, R. Stinshoff, Y. Zhang, J. Gooth, N. Kumar, G. Kreiner, K. Koepernik, R. Car, J. Kübler, G. H. Fecher, C. Shekhar, Y. Sun, and C. Felser, From Colossal to Zero: Controlling the Anomalous Hall Effect in Magnetic Heusler Compounds via Berry Curvature Design. Physical Review X **8**, 041045 (2018).
[46]     J. Noky, Q. Xu, C. Felser, and Y. Sun, Large anomalous Hall and Nernst effects from nodal line symmetry breaking in $Fe_2MnX$ (X = P, As, Sb). Phys. Rev. B **99**, 165117 (2019).
[47]     See Supplementary Materials at xxx for details of synthesis, transport measurements, computations and relevant refereces[60-70].
[48]     A. A. Burkov, Giant planar Hall effect in topological metals. Phys. Rev. B **96**, 041110 (2017).
[49]     P. Li, C. H. Zhang, J. W. Zhang, Y. Wen, and X. X. Zhang, Giant planar Hall effect in the Dirac semimetal $ZrTe_5$. Phys. Rev. B **98**, 121108 (2018).
[50]     H. Li, H.-W. Wang, H. He, J. Wang, and S.-Q. Shen, Giant anisotropic magnetoresistance and planar Hall effect in the Dirac semimetal $Cd_3As_2$. Phys. Rev. B **97**, 201110 (2018).
[51]     N. Kumar, S. N. Guin, C. Felser, and C. Shekhar, Planar Hall effect in the Weyl semimetal GdPtBi. Phys. Rev. B **98**, 041103 (2018).

[52] Y. J. Wang, J. X. Gong, D. D. Liang, M. Ge, J. R. Wang, W. K. Zhu, and C. J. Zhang, Planar Hall effect in type-II Weyl semimetal $WTe_2$. arXiv:1801.05929.
[53] F. C. Chen, X. Luo, J. Yan, Y. Sun, H. Y. Lv, W. J. Lu, C. Y. Xi, P. Tong, Z. G. Sheng, X. B. Zhu, W. H. Song, and Y. P. Sun, Planar Hall effect in the type-II Weyl semimetal $T_d$-$MoTe_2$. Phys. Rev. B **98**, 041114 (2018).
[54] R. Singha, S. Roy, A. Pariari, B. Satpati, and P. Mandal, Planar Hall effect in the type-II Dirac semimetal $VAl_3$. Phys. Rev. B **98**, 081103 (2018).
[55] S. Chadov, X. Qi, J. Kübler, G. H. Fecher, C. Felser, and S. C. Zhang, Tunable multifunctional topological insulators in ternary Heusler compounds. Nat. Mat. **9**, 541 (2010).
[56] H. Lin, L. A. Wray, Y. Xia, S. Xu, S. Jia, R. J. Cava, A. Bansil, and M. Z. Hasan, Half-Heusler ternary compounds as new multifunctional experimental platforms for topological quantum phenomena. Nat. Mat. **9**, 546 (2010).
[57] D. Xiao, Y. Yao, W. Feng, J. Wen, W. Zhu, X.-Q. Chen, G. M. Stocks, and Z. Zhang, Half-Heusler Compounds as a New Class of Three-Dimensional Topological Insulators. Phys. Rev. Lett **105**, 096404 (2010).
[58] B. G. Ueland, A. Kreyssig, K. Prokeš, J. W. Lynn, L. W. Harriger, D. K. Pratt, D. K. Singh, T. W. Heitmann, S. Sauerbrei, S. M. Saunders, E. D. Mun, S. L. Bud'ko, R. J. McQueeney, P. C. Canfield, and A. I. Goldman, Fragile antiferromagnetism in the heavy-fermion compound YbBiPt. Phys. Rev. B **89**, 180403 (2014).
[59] K.-Y. Yang, Y.-M. Lu, and Y. Ran, Quantum Hall effects in a Weyl semimetal: Possible application in pyrochlore iridates. Phys. Rev. B **84**, 075129 (2011).
[60] C. Liu, Y. Lee, T. Kondo, E. D. Mun, M. Caudle, B. N. Harmon, S. L. Bud'ko, P. C. Canfield, and A. Kaminski, Metallic surface electronic state in half-Heusler compounds RPtBi (R=Lu, Dy, Gd). Phys. Rev. B **83**, 205133 (2011).
[61] P. C. Canfield, J. D. Thompson, W. P. Beyermann, A. Lacerda, M. F. Hundley, E. Peterson, Z. Fisk, and H. R. Ott, Magnetism and heavy fermion‐like behavior in the RBiPt series. J. Appl. Phys. **70**, 5800 (1991).
[62] P. C. Canfield and Z. Fisk, Growth of single crystals from metallic fluxes. Philosophical Magazine B **65**, 1117 (1992).
[63] P. Hohenberg and W. Kohn, Inhomogeneous Electron Gas. Phys. Rev. **136**, B864 (1964).
[64] G. Kresse and J. Furthmüller, Efficient iterative schemes for *ab initio* total-energy calculations using a plane-wave basis set. Phys. Rev. B **54**, 11169 (1996).
[65] G. Kresse and D. Joubert, From ultrasoft pseudopotentials to the projector augmented-wave method. Phys. Rev. B **59**, 1758 (1999).
[66] J. P. Perdew, K. Burke, and M. Ernzerhof, Generalized Gradient Approximation Made Simple. Phys. Rev. Lett **77**, 3865 (1996).
[67] M. G. Haase, T. Schmidt, C. G. Richter, H. Block, and W. Jeitschko, Equiatomic Rare Earth (Ln) Transition Metal Antimonides LnTSb (T=Rh, lr) and Bismuthides LnTBi (T=Rh, Ni, Pd, Pt). J. Solid State Chem. **168**, 18 (2002).
[68] O. Pavlosiuk, X. Fabreges, A. Gukasov, M. Meven, D. Kaczorowski, and P. Wiśniewski, Magnetic structures of REPdBi half-Heusler bismuthides (RE = Gd, Tb, Dy, Ho, Er). Physica B **536**, 56 (2018).
[69] N. Marzari and D. Vanderbilt, Maximally localized generalized Wannier functions for composite energy bands. Phys. Rev. B **56**, 12847 (1997).

[70]  Q. Wu, S. Zhang, H.-F. Song, M. Troyer, and A. A. Soluyanov, WannierTools: An open-source software package for novel topological materials. Computer Physics Communications **224**, 405 (2018).

# Supplementary Materials for

# Exceptionally large anomalous Hall effect due to anticrossing of spin-split bands in the antiferromagnetic half-Heusler compound TbPtBi

**Details of Synthesis, transport measurements and Computations**

Single crystals of TbPtBi were synthesized using the Bi-flux method [1-3]. The starting materials of Tb, Pt and Bi powder were mixed with a molar ratio of Tb:Pt:Bi=1:1:20, loaded into an $Al_2O_3$ crucible, and sealed in a quartz tube under high vacuum. The mixture was heated to 1050 °C in a crucible furnace and held at this temperature for 48 hours for homogeneous melting, followed by slow cooling down to 700 °C at a rate of 3 C°/h. Cubic-shaped single crystals of TbPtBi were obtained by removing the excess Bi flux by centrifuging. Cubic structures of the crystals were confirmed by X-ray diffraction measurements. Magnetotransport measurements were made using a standard four-probe method in a Physical Property Measurement System (PPMS, Quantum Design). High-field measurements were carried out at the National High Magnetic Field Laboratory (NHMFL) in Tallahassee. All the samples used for transport measurements were polished to rectangular shapes, with the polished surfaces being parallel to the (100) plane. The current was applied in the (100) plane along the [010] or [001] direction, for both longitudinal ($\rho_{xx}$) and transverse (Hall) resistivity ($\rho_{xy}$) measurements. Magnetization measurements were performed with a SQUID magnetometer (Quantum Design).

First-principles calculations on TbPtBi were performed within the density-functional theory (DFT) framework using the Vienna ab-initio simulation package (VASP)[4,5] . Generalized gradient approximation (GGA) with Perdew-Burke-Ernzerhof parametrization was used to include exchange-correlation effects. [6,7] An on-site Coulomb interaction was added for Tb $f$-electrons within the $GGA+U$ scheme with $U_{eff}$ = 10 eV. An energy cut-off of 500 eV was used for the plane-wave basis set, and an 11 × 11 × 11 Γ-centered $k$ mesh was employed to sample the primitive bulk Brillouin zone (BZ). We employed experimental lattice parameters [8,9] with spin-orbit coupling effects included self-consistently in our computations. Topological properties were determined by employing a tight-binding model using the WannierTools package. We considered $d$ orbitals for Tb, $s$, $p$, and $d$ orbitals for Pt, and $p$ orbitals for Bi to construct the Wannier functions within the VASP2WANNIER90 interface [10,11].

**Supplementary Figure S1** | The Curie-Weiss fitting of the inverse magnetic susceptibility in the temperature range of 60-300K.

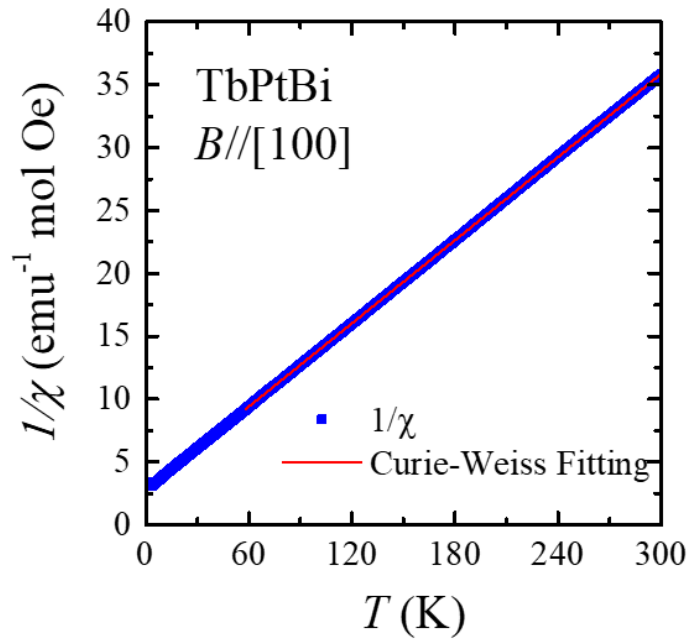

**Supplementary Figure S2** | (a) The measured anomalous Hall conductivity from Sample 1 at various temperatures. (b) Anomalous Hall angle as a function of the magnetic field $B$ for Sample 2 (S2).

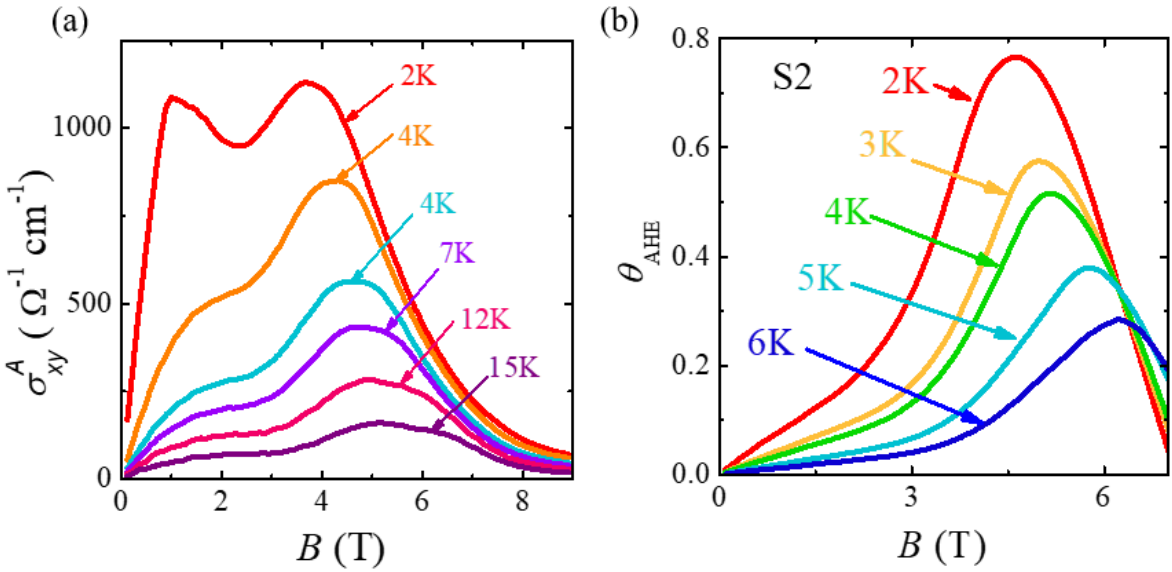

**Supplementary Figure S3** | (a) Magnetoresistivity of TbPtBi (S3) measured under high magnetic field ($B$ up to 35T) when $B//I$ ($B$ is along [100]). The SdH quantum oscillations start to develop when magnetic field $B$ is increased above 10T. (b) FFT spectra of the SdH oscillations. The Fermi wave vector extracted from the oscillation frequencies (104T, 134T) are 0.0562 and 0.0638 Å$^{-1}$, respectively.

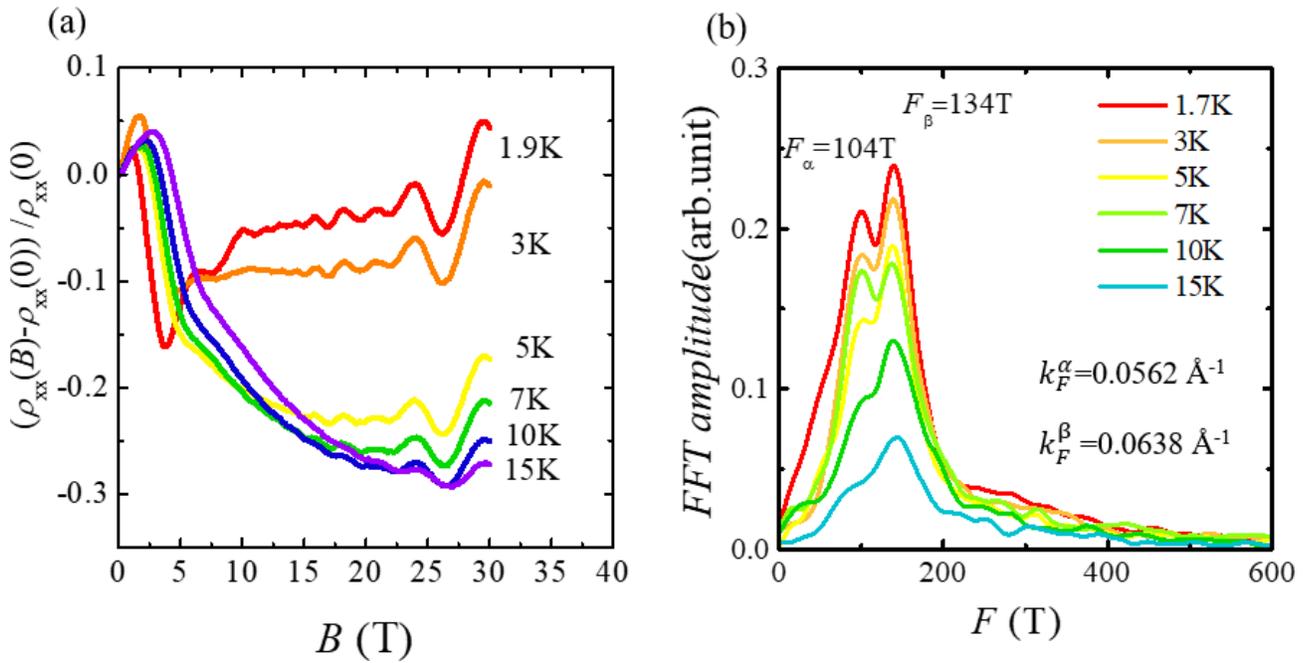

**Supplementary Figure S4** | (a) The anomalous Hall conductivity and (b) anomalous Hall angle estimated by subtracting the $\sigma_{xy}(50K)$ as ordinary Hall terms from Sample 1 at various temperatures.

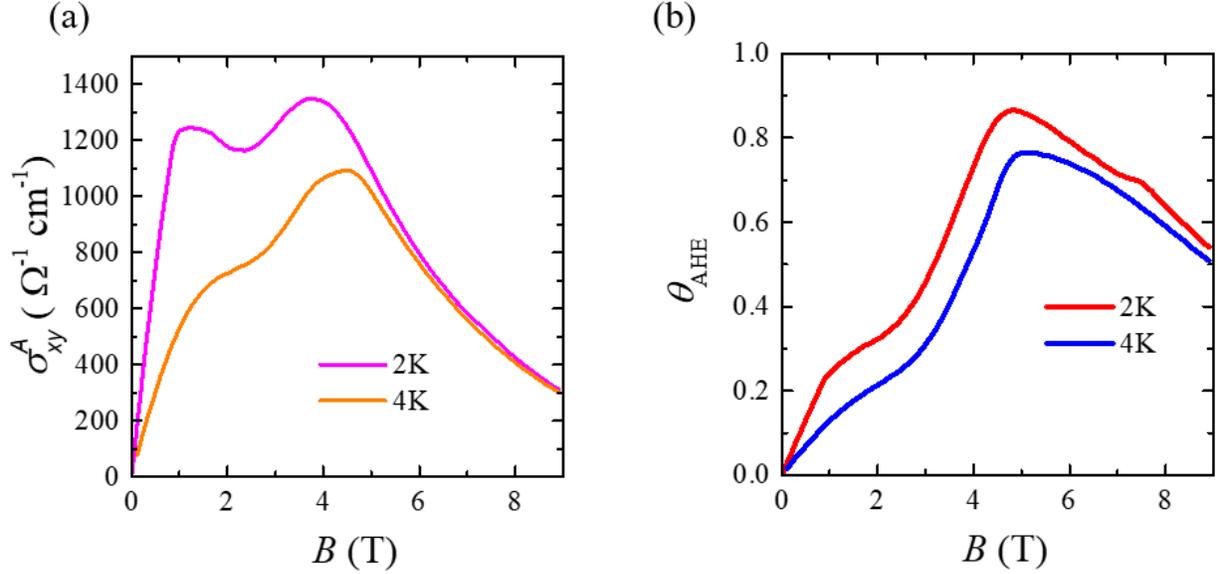

**Table S1** | Momentum-space location of $W^1_{100}$ Weyl nodes at 83 meV in the first-Brillouin zone with magnetization along the [100] direction in units of 1/Å.

| Weyl nodes | $k_x$ | $k_y$ | $k_z$ | Chirality | Energy |
|---|---|---|---|---|---|
| $W^1_{100}$ | 0.077 | -0.002 | 0.020 | +1 | 83 meV |
| | 0.081 | 0.020 | -0.030 | +1 | |
| | -0.077 | -0.021 | 0.003 | -1 | |
| | -0.081 | 0.030 | -0.020 | -1 | |

## References

[1]   C. Liu, Y. Lee, T. Kondo, E. D. Mun, M. Caudle, B. N. Harmon, S. L. Bud'ko, P. C. Canfield, and A. Kaminski, Metallic surface electronic state in half-Heusler compounds RPtBi (R=Lu, Dy, Gd). Phys. Rev. B **83**, 205133 (2011).
[2]   P. C. Canfield, J. D. Thompson, W. P. Beyermann, A. Lacerda, M. F. Hundley, E. Peterson, Z. Fisk, and H. R. Ott, Magnetism and heavy fermion-like behavior in the RBiPt series. J. Appl. Phys. **70**, 5800 (1991).

[3]     P. C. Canfield and Z. Fisk, Growth of single crystals from metallic fluxes. Philosophical Magazine B **65**, 1117 (1992).
[4]     P. Hohenberg and W. Kohn, Inhomogeneous Electron Gas. Phys. Rev. **136**, B864 (1964).
[5]     G. Kresse and J. Furthmüller, Efficient iterative schemes for *ab initio* total-energy calculations using a plane-wave basis set. Phys. Rev. B **54**, 11169 (1996).
[6]     G. Kresse and D. Joubert, From ultrasoft pseudopotentials to the projector augmented-wave method. Phys. Rev. B **59**, 1758 (1999).
[7]     J. P. Perdew, K. Burke, and M. Ernzerhof, Generalized Gradient Approximation Made Simple. Phys. Rev. Lett **77**, 3865 (1996).
[8]     M. G. Haase, T. Schmidt, C. G. Richter, H. Block, and W. Jeitschko, Equiatomic Rare Earth (Ln) Transition Metal Antimonides LnTSb (T=Rh, Ir) and Bismuthides LnTBi (T=Rh, Ni, Pd, Pt). J. Solid State Chem. **168**, 18 (2002).
[9]     O. Pavlosiuk, X. Fabreges, A. Gukasov, M. Meven, D. Kaczorowski, and P. Wiśniewski, Magnetic structures of REPdBi half-Heusler bismuthides (RE = Gd, Tb, Dy, Ho, Er). Physica B **536**, 56 (2018).
[10]    N. Marzari and D. Vanderbilt, Maximally localized generalized Wannier functions for composite energy bands. Phys. Rev. B **56**, 12847 (1997).
[11]    Q. Wu, S. Zhang, H.-F. Song, M. Troyer, and A. A. Soluyanov, WannierTools: An open-source software package for novel topological materials. Computer Physics Communications **224**, 405 (2018).